 \documentclass [twocolumn,ashowpacs,preprintnumbers,amsmath,amssymb]{revtex4}
\usepackage{hyperref}

\usepackage{graphicx}

\usepackage{dcolumn}
\usepackage{amssymb}
\newcommand{\be}{\begin{eqnarray}}
\newcommand{\ee}{\end{eqnarray}}

\newcommand{\pr}[1]{Phys. Rev. {\bf #1}}

\newcommand{\np}[1]{Nucl. Phys. {\bf #1}}

\begin{document}

\title {All orders proton breakup from exotic nuclei}

\author{ A. Garc\'ia-Camacho$^{(a)}$, G. Blanchon$^{(a)}$, A. Bonaccorso$^{(a)}$, D. M. Brink$^{(b)}$  \\ \small
     $^{(a)}$ INFN, Sez. di Pisa and Dipartimento di Fisica, Universit\`a di Pisa,\\\small Largo Pontecorvo 3, 56127
Pisa, Italy.\\ \small 
  $^{(b)}$Department of Theoretical Physics, 1 Keble Road, Oxford OX1 3NP, U. K.\\
 }

\begin{abstract}
 We present a semiclassical method  to treat the proton breakup from a weakly bound state in an exotic nucleus. The Coulomb interactions between the proton, core and  target are treated to all orders and including the full multipole expansion of the Coulomb potential. The nuclear proton-target interaction is also treated to all orders. The core-target interaction is included as an absorption. The method is semi-analytical thus allowing for a detailed understanding of the short range and long range effects of the interactions in the reaction dynamics. It  explains also the origin of possible asymmetries in the core parallel momentum distributions when the full multipole expansion of the Coulomb potential is used. Calculations are compared to results of other, fully numerical, methods and to experimental data in order to establish the accuracy and reliability of the method.
\end{abstract}

\maketitle
{\bf Pacs} {21.10.Jx, 24.10.-i, 25.60.Gc, 27.30.+t}
\section{Introduction}

 The study of exotic nuclei through knockout reactions has provided important results along the last decades. Among those, one proton removal reactions have been proven as a useful tool for the understanding of light, proton rich, exotic nuclei. From the reaction point of view, the inclusion of a long-range part in the interaction between the proton and the target, i.e. the Coulomb field, provides a significant complication with respect to the neutron case as well as a challenge for reaction theory.

 This work is an extension of the method presented in Ref. \cite{nois}. There, a formalism to treat neutron breakup to all orders, including the whole multipole expansion of the Coulomb interaction giving the recoil of the core, was presented. Here that theoretical framework is completed in order to account for proton breakup as well. 
 For proton halo nuclei Coulomb breakup reactions in the
laboratory have been used to get indirect information on the radiative capture, since it has been
shown that the Coulomb breakup cross section is proportional to
the radiative capture cross section \cite{carlos1}. A number of experimental papers \cite{negoita96}-\cite{borcea} have reported experiments on proton breakup from $^8$B. $^{8}$B is  partner in (p,$\gamma$) radiative
capture reactions of great astrophysical interest for the
understanding of the neutrino flux from the sun (see for example
the discussion and references of \cite{davids01}).  $^{17}$F is another candidate for proton halo which is still  under investigation both experimentally \cite{rem,dwb1,flo17f} as well as theoretically \cite{esb01,esbber1,esben96,carlos03}. Furthermore $^{17}$F plays an important role in understanding 
explosive nucleo-synthesis in X-ray bursts and novae, as it enters
in the sequence $^{14}$O($\alpha$,p)$^{17}$F(p,$\gamma$)$^{18}$Ne($\alpha$,p)$^{21}$Na, where
cross sections are poorly known.

The existence of
 a proton halo has sometimes been questioned \cite{kelly} and
results from different experiments might seem to be contradictory \cite{gai,ang04}.  Some of the apparent discrepancies in experimental results and their analysis have been recently discussed and in  large part solved in Ref. \cite{esben05}.
  \section{Formalism} \label{teo}
\begin{figure}
\center
\includegraphics[scale=0.1,width=5cm]{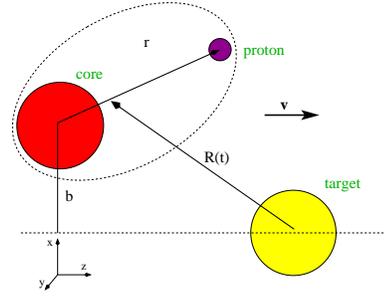}
\caption{(Colour online) Coordinate system.}
\label{pall}
\end{figure}
 This work is the extension to the proton breakup problem of the formalism developed in \cite{nois} for neutron breakup. Similar formulae to those of the neutron breakup can be derived, since the Coulomb potential can be written as
\begin{eqnarray}
V(\vec{r},\vec{R})= \frac{V_c}{|\vec{R}-\beta_1 \vec{r}|}+\frac{V_v}{|\vec{R}+\beta_2 \vec{r}|}-\frac{V_0}{R}
\end{eqnarray}
 where $V_c=Z_cZ_te^2$, $V_v=Z_vZ_te^2$ and $V_0=(Z_v+Z_c)Z_te^2$. $\beta_1$ and $\beta_2$ are the mass ratios of proton and core, respectively, to that of the projectile. The coordinate system used in this paper is shown in Fig. \ref{pall}. Following a procedure analogous to that of \cite{nois}, where $\chi$ was defined  as $\chi_{pert}=\frac{1}{\hbar}\int dt e^{i\omega t} V(\vec{r},t)$, the Coulomb phase for the proton is shown to be
\begin{eqnarray} \label{fasipr}
\chi^p&=&\frac{2}{\hbar v} (V_c e^{i\beta_1 \omega z /v}K_0(\omega b_c/v) -V_0 K_0(\omega R_\perp/v) \nonumber  \\ &&+V_v e^{-i\beta_2 \omega z /v}K_0(\omega b_v/v)  )
\end{eqnarray}
with $\omega=\left(  {{\varepsilon  _f}}-\varepsilon_{0}\right)/\hbar
$ and $\varepsilon_0$ is the neutron initial bound state energy  while
${{\varepsilon  _f}}$ is the  final neutron-core continuum energy.
 Since $V_0=V_c+V_v$, Eq. (\ref{fasipr}) can be written as
\begin{eqnarray} \label{pff}
\chi^p=\chi(\beta_1,V_c)+\chi(-\beta_2,V_v)
\end{eqnarray}
   leading to
\begin{eqnarray} \label{fiqui}
\chi(\beta,V)=\frac{2 V}{\hbar v}\left(e^{i\beta \omega z /v}K_0(\omega b/v) -K_0(\omega R_\perp/v)\right).
\end{eqnarray}

 The Coulomb phase Eq.(\ref{pff}) is therefore the sum of two terms: one of them describes the recoil of the core whereas the other accounts for the direct proton-target Coulomb interaction. Of course, in the case of the neutron the latter vanishes and the phase reduces to the one derived in \cite{nois}.

 The expansion of Eq. (\ref{fiqui}) to first order in $\vec{r}$ yields the dipole approximation to the phase:
\begin{eqnarray}\label{dipp}
\chi^p &\simeq& \frac{2 (\beta_1V_c-\beta_2V_v)}{\hbar v}(K_0(\omega R_\perp/v)\frac{i\omega z}{v} \nonumber \\ &&+K_1(\omega R_\perp/v)\frac{\vec{R}_\perp \cdot \vec{r}}{R_\perp}\frac{\omega}{v}),
\end{eqnarray}
 which only differs from the neutron breakup case in the different constant factor, which is now $(V_c\beta_1-V_v\beta_2)$ instead of $V_0\beta_1$ of Ref. \cite{nois}.

Our expression for the differential cross-section is
\begin{eqnarray}
\frac{d \sigma}{d \vec{k}}=\frac{1}{8\pi^3}\int d \vec{b}_c |S_{ct}(b_c)|^2 |g^{rec}+g^{dir}+g^{nuc}|^2.
\label{cross}\end{eqnarray}
where $|S_{ct}(b_c)|^2$ is the core-target elastic scattering probability discussed later. Such a form of the cross section has been introduced for relativistic  energy electromagnetic excitations in Ref.\cite{aw} where however a first order perturbation theory amplitude was considered. It has  been derived in an eikonal approach to the halo-projectile scattering \cite{mar02,mar03} and it  has been widely used  also for normal  projectiles in the case of  peripheral reactions such as transfer and inelastic excitations \cite{bw}. Under the conditions that the core-target interaction is mainly absorptive it has been re-obtained by Bertulani in a  relativistic approach to proton breakup \cite{bertu}. 
This paper has also shown that relativistic corrections to  dynamical approaches such as the continuum discretised coupled channel method (CDCC) can  increase the proton breakup cross section  of as much as 15\% at the incident energy of 250 A.MeV. Our formalism however follows closely Ref.\cite{aw} where relativistic kinematics is easily implemented and it has been used in the numerical calculations presented in this paper.  As we shall see in the following our results agree indeed rather well with high energy experimental data. 

The probability amplitude in Eq.(\ref{cross}) has been written as the sum of three pieces:
the recoil term,
\begin{eqnarray}
g^{rec}&=&\int d\vec{r} e^{-i \vec{k} \cdot \vec{r}} \phi_i(\vec{r}) ( e^{i \frac{2 V_c}{\hbar v}\log{\frac{b_c}{R_\perp}}} -1 -i\frac{2 V_c}{\hbar v}\log{\frac{b_c}{R_\perp}} \nonumber \\ && +i\chi(\beta_1,V_c) ),
\label{grec}\end{eqnarray}
 where, according to the discussions in \cite{nois,typ01a,mar02,mar03}, the sudden limit has been used in order to include all orders in the interaction. Similarly, the second term in our probability amplitude is the direct proton Coulomb interaction. It has the same form as Eq.(\ref{grec}) but for the substitution $V_c\to V_v$, $b_c \to b_v$ and $\beta_1 \to -\beta_2$.
 
 Finally, the nuclear part is 
\begin{eqnarray}
g^{nuc}=\int d\vec{r} e^{-i \vec{k} \cdot \vec{r}} \phi_i(\vec{r}) \left( e^{i \chi_{nt}(b_v)} -1 \right).
\end{eqnarray}
 
  A number of papers have addressed the problem of asymmetry in the core parallel momentum distribution after proton knockout \cite{davids98,davids01,davids03,esben96,esben00,esben95,tt}. The fact that this asymmetry comes from high order terms can be directly extracted from our formalism. If the Coulomb part of the amplitude $g^{Cou}=g^{rec}+g^{dir}$ is simply expanded to first order in $\chi$, it becomes
\begin{eqnarray}
g^{Cou}&\simeq &\int d\vec{r} e^{-i \vec{k} \cdot \vec{r}} \phi_i(\vec{r})\frac{2}{\hbar v}(V_c e^{i\beta_1 \omega z /v}K_0(\omega b_c/v) \nonumber \\ &&-V_0 K_0(\omega R_\perp/v) 
 +V_v e^{-i\beta_2 \omega z /v}K_0(\omega b_v/v)),
\end{eqnarray}
 which can be written, in terms of the one-dimensional Fourier transform in $z-$direction $\hat{\phi}_i$, as
\begin{eqnarray}
g^{Cou}&\simeq &\int d\vec{r}_\perp e^{i \vec{k}_\perp \cdot \vec{r}_\perp} \frac{2}{\hbar v}\left(V_c K_0(\omega b_c/v)\hat{\phi}_i(\vec{r}_\perp,k_z+\beta_1 \omega/v )\right. \nonumber \\ && -\left.V_0 K_0(\omega R_\perp/v)\hat{\phi}_i(\vec{r}_\perp,k_z) \right. \nonumber \\ &&
 +\left.V_v K_0(\omega b_v/v)\hat{\phi}_i(\vec{r}_\perp,k_z-\beta_2 \omega/v)\right)\label{11}
\end{eqnarray}
 Thus the Coulomb breakup probability amplitude can be regarded as a coherent sum of three terms, each of which contains a shifted $z-$Fourier transform. The shifts are in opposite directions, $-\beta_1 \omega/v$ and $\beta_2 \omega/v$, but they are not visible  in the calculated momentum distributions shown in this paper as $\omega$ depends on $k$ itself. The amount of deformation will also depend on the explicit form of the wave function ${\phi}_i(\vec{r})$. Moreover, the $1/v$ factor indicates that the asymmetry decreases as the beam energy increases.

 In dipole approximation, however, the amplitude is just

\begin{eqnarray}
g^{Cou}&\simeq &\int d\vec{r}_\perp e^{-i \vec{k}_\perp \cdot \vec{r}_\perp}\frac{2 (\beta_1V_c-\beta_2V_v)}{\hbar v} \nonumber \\
&& \times\left(K_0(\omega R_\perp/v)\frac{\omega}{v}\frac{d}{d k_z}\hat{\phi}_i(\vec{r}_\perp,k_z)  \right.\nonumber \\
&&\left.+K_1(\omega R_\perp/v)\frac{\vec{R}_\perp \cdot \vec{r}}{R_\perp}\frac{\omega}{v}\hat{\phi}_i(\vec{r}_\perp,k_z)\right),
\end{eqnarray}
 which does not contain any asymmetry for the momentum distribution as it involves square moduli of $\hat{\phi}_i(\vec{r}_\perp,k_z)$ separately. Hence we can confirm analytically that the asymmetry in Coulomb breakup parallel momentum distributions is due to the presence of higher multipole terms, in agreement with earlier works \cite{davids98,davids01,davids03,esben96,esben00,esben95}. However, the presence of the nuclear interaction introduces an interference that does depend on the sign of $k_z$ and thus an additional asymmetry to that due to higher multipole terms in the Coulomb interaction.

\section{Applications}
\begin{figure}
\center
\includegraphics[scale=0.2,width=6.cm]{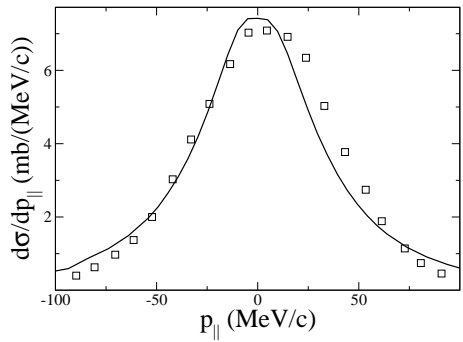}
\caption{Calculated inclusive momentum distribution of $^7$Be fragments after proton-removal from $^8$B against Pb at 936 A.MeV. The calculations are the sum of contributions leading to the ground and excited state of $^7$Be weighted by the respective spectroscopic factors and  include both the diffraction and stripping parts of nuclear breakup, besides Coulomb breakup. Data are from \cite{lola03}.}
\label{ggg}
\end{figure}

\begin{figure}
\center
\vskip 15pt
\includegraphics[scale=0.2,width=7cm]{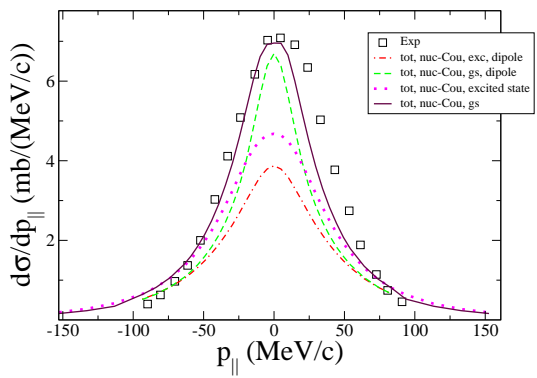}
\caption {(Colour online) Proton momentum distribution after Coulomb breakup of  $^8$B against Pb at 936 A.MeV in both dipole and full-multipole approximations, for both ground and excited state. Spectroscopic factors are not included.}
\label{gsi}
\end{figure}
 This formalism has been applied to proton breakup of $^8$B against Pb and C targets at a beam energy of 936 A.MeV, experiment reported in \cite{lola03}. The  projectile $^8$B is taken  as a two-body object. Its ground state has components with a proton coupled to the ground $3/2^{-}$ state   and to the excited E$^* =429$keV $1/2^{-}$ state of $^7$Be. The proton separation energy is 137 keV when the $^7$Be is left in its ground state and $137 + 429$ keV when it is left in an excited state.

Radial wave functions have been obtained by numerical solution of the Schr\"odinger equation in Woods-Saxon potentials with depths adjusted to reproduce the corresponding separation energies. The radius parameter of these Woods-Saxon potentials have been taken as 1.3 fm with a diffuseness of 0.6 fm.

 The nuclear breakup is described in this formalism through the proton-target and core-target S-matrices. These have been calculated by taking the optical limit in Glauber theory \cite{jeff99}. The effective two-body interactions are obtained by folding effective nucleon-nucleon interactions with the densities of the nuclei involved. Pb density was approximated by a Woods-Saxon profile fitted to reproduce the experimental root mean square radius 5.5 fm \cite{kara02}. $^7$Be and $^{12}$C densities were taken as gaussian with root mean square radius 2.31 and 2.35 fm respectively \cite{oza01}. The stripping contribution to nuclear breakup has been calculated in the eikonal formalism \cite{hen96}.

With these ingredients and the formulae described in the previous section, the single particle cross-sections of Table \ref{t1} were calculated. Interference between Coulomb and elastic nuclear part are included automatically in the amplitude, but stripping must be added incoherently. When comparing to experimental data, spectroscopic factors must be taken into account. Following \cite{lola03}, the spectroscopic factor for the excited state is $S_{ex}=0.101$ in this work, while for the ground state we have used $S_{gs}=0.841$. The latter has been obtained by adding together the spectroscopic factors for two states which differ in core intrinsic spin \cite{lola03}, since our formalism cannot distinguish between them. These numbers provide an exclusive/inclusive ratio of $R_{theo}=0.089$ for the Pb target, in good agreement with the experimental result $R_{exp}=0.085 \pm 0.021$. Such agreement is also good for the C target, as the experimental result is $R_{exp}=0.13 \pm 0.03$, while we obtain $R_{theo}=0.098$. The inclusive momentum distribution on lead is shown in Fig. \ref{ggg}  with no scale factors. In this calculation the stripping contribution is assumed to have the same shape as the diffractive one. Clearly the agreement is not perfect, but in view of the very simple two-body forms of our initial and final wave functions which do not contain any three-body effects, it can be considered reasonable. Regarding absolute cross-sections, in the case of the Pb target, our result for the excited state is 57.8 mb, in agreement with the exclusive measurement of $56 \pm 5$ mb, while the inclusive result $662 \pm 60$ mb is rightly described by our calculation of 652.4 mb (Table \ref{t1}). For the C target we predict 9.1 (92.6) mb for the exclusive (inclusive) measurement, consistent with the experimental result $12 \pm 3$ ($94 \pm 9$) mb.
 
 This formalism allows a direct check of the dipole approximation for the Coulomb interaction, by comparing the results provided by the phase Eq.(\ref{pff}) to those of Eq.(\ref{dipp}). Fig. \ref{gsi}  shows the calculated momentum distributions of protons in the projectile reference frame after breakup of $^8$B. This does not include stripping. The dipole approximation, yields narrower momentum distribution and smaller single particle cross-section. High order effects, especially in the heavy target case, are not negligible, in agreement with \cite{capel05}, although in their calculation the effect appears smaller due to their choice of large impact parameters.

We have also performed preliminary calculations for  $^{17}$F. They differ from those of Ref.\cite{mb2}
by about an order of magnitude for the g.s. breakup and by about a factor three for the breakup from 
$^{17}$F$^*$.
The reason for this discrepancy is in large part due to the use of a the plane wave as  final state which neglects the Coulomb proton-core interaction. An estimate of the possible correction can be obtained by calculating a barrier penetration factor $w$ at  the core radius R$_c$  according to Ref.\cite{LL}.

In Table \ref{t2} we give the values of the penetration factors $w_1$ and $w_2$ for   $^8$B and $^{17}$F projectiles respectively.  We give also the values of the corresponding Sommerfeld parameters n$_1$ and n$_2$, with n=Z$_p$Z$_t$e$^2/\hbar v$. These values show that for a peak in the proton-core energy spectrum around 1.5-2MeV there is a possible error in the cross sections of 10-20\% in the $^8$B projectile case while the error is of 40-60\% for $^{17}$F. Such estimates are independent from the initial binding energy and from the beam energy. Our method is an high energy approximation
and therefore further corrections should be expected if the method is applied at the low energies considered in Ref.\cite{mb2}

\begin{table}
\begin{center}
\small
\caption{ Single particle cross-sections in mb. Beam energies in A.MeV.}
\medskip
\begin{tabular}{ccccccc }
\hline\hline Projectile & Target~~~&Energy&~~ $\sigma_{Cou+diff}$ ~~& ~~$\sigma_{st}$~~&~~$\sigma_{sp}$~~& S$\times \sigma_{sp}$\\ \hline
$^8$B & Pb & 936 & 558 & 149 & 707 & 594.6 \\
$^8$B$^*$ & Pb & 936 & 442 & 130 & 572 & 57.8\\ 
$^8$B  & C  & 936 & 36.9& 62.4& 99.3 & 83.5 \\
$^8$B$^*$ & C  & 936 & 32.1& 57.4& 89.5 & 9.1\\
  \hline \hline\label{t1}
\end{tabular} 
\\
\end{center}\end{table}

\begin{table}
\begin{center}
\small
\caption{Barrier penetration factors $w_1$ and $w_2$ for $^8$B and $^{17}$F respectively.  Proton-core energies, $\varepsilon_f$, and momenta, $ \hbar k$, in MeV. n$_1$ and n$_2$ are dimensionless Sommerfeld parameters.}
\medskip
\begin{tabular}{cccccc}
\hline\hline 
$\varepsilon_f$ &$ \hbar k$ &n$_1$ &$w_1$&n$_2$&$w_2$ \\ \hline
  0.5& 30.7&0.89&0.22&1.78&0.013\\ 
  1.0&43.5&0.63&0.58&1.26&0.158\\ 
  1.5& 53.2&0.51&0.81&1.03&0.39\\ 
  2.0&61.5&0.44&0.94&0.89&0.61\\ 
    \hline \hline
\label{t2}\end{tabular} 
\\
\end{center}\end{table}
 \section{Conclusions}

In this paper we have introduced a semiclassical and analytical  method to treat proton breakup to all orders and all multipoles in the Coulomb potential. As previously done by some of us \cite{nois,mar02,mar03}, the present method is based on the use of a final nucleon wave function which is  an extention of the traditional eikonal-type. If the Coulomb potential is neglected, our breakup amplitude reduces to the well known eikonal form for the nuclear breakup. On the other hand if only the first order term of the Coulomb phase is considered, our Coulomb amplitude reduces to the  multipole expansion in which the dipole term is treated by perturbation theory, while high order  and higher multipole terms are treated in the sudden approximation according to \cite{mar03}.
To test the method we have applied it to the study of one proton breakup from both the ground and first excited states of $^8$B.  Our results are in very good agreement with experimental data at relativistic energies. Preliminary calculations for $^{17}$F at lower energies (65 A.MeV) have shown discrepancies of about one order of magnitude with respect to the results of Ref.\cite{mb2}. We believe this is mainly due to our choice of  a plane-wave for the  final state wave-function.  However, the use of the plane-wave has allowed analytical calculations leading to Eq.(\ref{11}) which is perhaps the most interesting result of this work. It   shows explicitly the origin of a possible asymmetry in core parallel momentum distributions as contained in  the Fourier transform of the initial state wave function.  In agreement with other numerical methods \cite{esben96} we have demonstrated that the asymmetry is due to the  higher order multipole terms in the expansion of the Coulomb potential, while it disappears in the dipole approximation.

\end{document}